\documentstyle[aps,epsfig]{revtex}

\begin{document}

\vspace{2.5cm}
\begin{center}
   {\Large \bf Measures of Charge Fluctuations

 in Nuclear Collisions   

   }  
\end{center}

\vspace{0.5cm}

\begin{center}

{\bf Jacek Zaranek}\footnote{E--mail: Jacek.Zaranek@cern.ch}\\
\vspace{0.3cm}
Institut f\"ur Kernphysik, Universit\"at Frankfurt\\
D--60486 Frankfurt, Germany\\[0.8cm]

\end{center}

\vspace{0.5cm}

\begin{abstract}
\noindent
The properties of two measures of charge fluctuations $\tilde{D}$ and $\Delta\Phi_q$ are discussed within several toy models of nuclear collisions. In particular their dependence on mean particle multiplicity, multiplicity fluctuations and net electric charge are studied. It is shown that the measure $\Delta\Phi_q$ is less sensitive to these trivial biasing effects than the originally proposed measure $\tilde{D}$. Furthermore the influence of resonance decay kinematics is analysed and it is shown that it is likely to shadow a possible reduction of fluctuations due to QGP creation.
\end{abstract}

\newpage
\section{Introduction}
Recently Jeon, Koch \cite{Ko.1} and Asakawa, Heinz and Mueller \cite{As.1} argued that the study of event-by-event fluctuations of electric charge in high energy nucleus-nucleus collisions may provide information on the state of matter in an early stage of the collision. The authors calculated the magnitude of the charge fluctuations in the Quark-Gluon Plasma (QGP) and in a hadron gas. These calculations show that the fluctuations in the QGP should be significantly smaller than in a hadron gas. Thus they can be used as a signal for the creation of the deconfined phase providing that the initial fluctuations survive hadronization and their relaxation time is significantly longer than the time of the hadronic stage of the collisions \cite{Ko.1,As.1,Sh.1}. 

The charge fluctuations should also be sensitive to other effects like the number of resonances at chemical freeze-out \cite{Ko.2,He.1} and fluctuations occurring in the initial stage \cite{Ma.3}. Consequently their analysis is interesting also beyond the QGP hypothesis. 

First experimental results on charge fluctuations in central collisions of heavy nucleus at SPS \cite{Bl.1} and RHIC \cite{To.1} are already available and don't show the expected reduction of fluctuations due to QGP creation.  

The analysis of charge fluctuations is relatively easy to perform experimentally due to the typically good resolution of the measurement of electric charge in tracking detectors positioned in a magnetic field.

The magnitude of the measured charge fluctuations is, however, also dependent on trivial, uninteresting effects, which may shadow the studied physics. The two most important of these effects are:

  - the fluctuations in the event multiplicity, introduced mostly by the variation of the impact parameter;

  - changes in the mean multiplicity due to changes of the colliding system (e.g. p+p, Pb+Pb), changes of collision energy and due to changes of the size of the acceptance in which fluctuations are studied. 

\noindent In order to minimize the sensitivity to these effects two measures of fluctuations were proposed:  $\tilde{D}$ \cite{Ko.3} and $\Phi$ \cite{Ma.1,Ma.2}.   

The aim of this paper is to study the properties of these measures using simple toy models of particle production in nuclear collisions.       
 
In Section II the fluctuation measures will be introduced. Their properties will be tested in Section III. Summary and conclusions are given in Section IV.

\section{MEASURES OF CHARGE FLUCTUATIONS: $\tilde{D}$ and $\Delta\Phi_{q}$}

The measure $\tilde{D}$ is defined as follows \cite{Ko.1,Ko.3} :

\begin{equation}\label{D}
\tilde{D} = \frac{\langle \delta R^2\rangle \cdot \langle N_{ch}\rangle}{C_y\cdot C_{\mu}} ,  
\end{equation}
where:
\begin{equation}\label{R}
R = \frac{N_{pos}}{N_{neg}} , 
\end{equation}
\begin{equation}\label{deltaR}
\delta R = R - \langle R\rangle , 
\end{equation}
\begin{equation}\label{Nch}
N_{ch} = N_{pos} + N_{neg} ,
\end{equation}
\begin{equation}\label{C}
C_{y} = 1 - \frac {\langle N_{ch}\rangle }{\langle N_{ch}\rangle _{tot}},\;\;\;\;\; C_{\mu} = \frac {\langle N_{pos}\rangle^2}{\langle N_{neg}\rangle^2} .  
\end{equation}

\noindent $N_{pos}$ and $N_{neg}$ are measured multiplicities of positively and negatively charged particles within the analysed acceptance. The symbol $\langle ...\rangle$ represents averaging over events. $\langle N_{ch}\rangle $ is the mean number of charged particles within the acceptance and $\langle N_{ch}\rangle _{tot}$ denotes the mean total event multiplicity of charged particles in full phase space.
The factors $C_y$ and $C_{\mu}$ were introduced to remove the influence of global charge conservation ($C_y$) and effect of non-zero net charge ($C_{\mu}$).   
Under the assumption of Poissonian distribution of $N_{pos}$ and $N_{neg}$ (providing the number of accepted particles is much smaller than the total number of particles $ \Rightarrow C_y\approx 1$) and assuming zero net charge ($Q=N_{pos,tot}-N_{neg,tot}=0 \Rightarrow C_{\mu}=1$) one obtains $\tilde{D}=4$ \cite{Ko.1,Ko.3}. Many hadrons measured in the final state originate from the decay of resonances. Therefore those are correlated in kinematic quantities like rapidity. Consequently if we consider an acceptance window in a kinematic variable which is large compared to the mean separation of the decay products in this variable the probability that all decay products will fall into it is high and  therefore the charge fluctuations should be reduced significantly \cite{Ko.1,Ko.2,Ko.3}. In the QGP phase the unit of charge is $\frac{1}{3}$ instead of $1$ as in the hadron phase. Smaller charge units cause smaller charge fluctuations. Assuming that the QGP fluctuations are frozen one expects $\tilde{D}\approx 1$ \cite{Ko.1,Ko.3}.

A well established measure of event-by-event fluctuations is the variable $\Phi$ \cite{Ma.1} which is defined in the following way:
\begin{equation}\label{Phi}
\Phi = 
\sqrt{\langle Z^2 \rangle \over \langle N \rangle } -
\sqrt{\overline{z^2}} \; ,
\end{equation} 

where:
\begin{equation}\label{z}
z = x - \overline{x},\;\;\;\;\;\;\;\;
Z = \sum_{i=1}^{N}(x_i - \overline{x}) . 
\end{equation}
$x$ is a single particle variable, $N$ is the event multiplicity within acceptance and overline denotes averaging over a single particle inclusive distribution. By construction, for a system which is an independent sum of identical particle sources the value of $\Phi$ is equal to the value of $\Phi$ for a single particle source independent of the number of superimposed particle sources and its distribution \cite{Ma.1}. In the original proposal \cite{Ma.1} $\Phi$ was introduced as a measure of fluctuations of a kinematic variable like transverse momentum. Furthermore it was proposed \cite{Ma.2} to use $\Phi$ as a measure of fluctuations of discrete quantities. Following this suggestion in this publication $x$ is identified with the particle charge $q$. Some properties of the $\Phi$ measure for discrete quantities were derived in Ref. \cite{Mr.1}. For a scenario in which particles are correlated only by global charge conservation ($gcc$) and providing $Q=0$ we calculated the value of $\Phi_q$:
\begin{equation}\label{Phiq1}
\Phi_{q,gcc}=\sqrt{1-P}-1 , 
\end{equation}
where
\begin{equation}\label{P}
P=\frac {\langle N_{ch}\rangle }{\langle N_{ch} \rangle _{tot}} .
\end{equation}
Thus the value of $\Phi_{q,gcc}$ depends only on the fraction of accepted particles. In order to remove the influence of $gcc$ we suggest therefore to use a difference:
\begin{equation}\label{deltaPhi}
\Delta \Phi_q =\Phi_q-\Phi_{q,gcc}\;.
\end{equation} 
 
Consequently the value of $\Delta \Phi_q$ is zero if the particles are correlated only by global charge conservation. It is negative in case of an additional correlation between positively and negatively charged particles and positive if there are anticorrelations. One possible source of additional correlations could be the creation of a QGP. The first part in the definition of $\Phi$ takes into account all, statistical and dynamical fluctuations. The second part quantifies only the statistical ones. Therefore the value of $\Phi$ calculated for different systems is independent of their degrees of freedom (quarks and gluons or hadrons) of the considered system if its fluctuations are only statistical. However, experimentally we measure the fluctuations of particles only after hadronization. A hadronization which conserves charge and entropy produces correlations. Consequently the measured value of $\Phi_q$ and $\Delta\Phi_q$ should be smaller, if a QGP was created in the early stage of the collision. 

\section{Toy Models of Nuclear Collisions}

In this section properties of the fluctuation measures, $\tilde{D}$ and $\Delta\Phi_q$, are discussed. First we will assume that the only source of correlations is the global charge conservation. Under this assumption we will study the behaviour of the measures at small $\langle N_{ch}\rangle$ and their response to non-zero net charge and fluctuations of event multiplicity. Furthermore we also calculate their values in two toy models, a '$\rho$'--gas model and a QGP model in which additional correlations are present. In the study we use a simple Monte Carlo (MC) procedure, which simulates the particle distribution in momentum space. The values of $\tilde{D}$ and $\Delta\Phi_q$ are calculated for particles falling into rapidity windows centred around mid-rapidity. In the MC simulations we used values of total multiplicity ($\langle N_{ch}\rangle _{tot}$) and net charge ($Q$) which are typical for central Pb + Pb collisions at 40 A$\cdot$GeV as measured by the NA49 collaboration \cite{Bl.1}. In the figures the measures are plotted as a function of $\langle N_{ch} \rangle$, which is a positive monotonic function of the size of the rapidity window. The ranges of the measures $\tilde{D}$ and $\Delta\Phi_q$ plotted on the vertical scales are chosen in order to make both measures similar sensitive to resonance decay kinematics and QGP creation. Statistical errors of the MC simulations are plotted but in most of the cases they are smaller than the symbol size.

\subsection{Model A}

In the first model we analyse the dependence of the measures $\tilde{D}$ and $\Delta\Phi_q$ on $\langle N_{ch}\rangle$. It is assumed that the total multiplicity is fixed, $N_{ch,tot}=850$, and the net charge in each event is zero ($Q=0$). The dependence of $\tilde{D}$ and $\Delta\Phi_q$ on $\langle N_{ch}\rangle$ is shown in Fig.~1.

\begin{figure}[p]
\epsfig{file=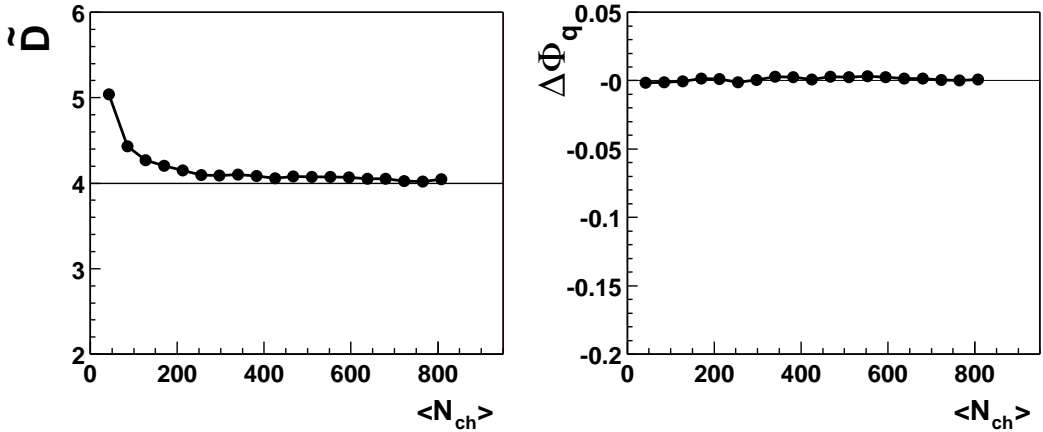,width=14cm}
\caption{
The dependence of $\tilde{D}$ and $\Delta\Phi_q$ on $\langle N_{ch}\rangle$ obtained in model A.  
}
\label{fig1}
\end{figure}

\noindent As expected the value of $\Delta\Phi_q$ is zero and the value of $\tilde{D}$ is close to 4 for large $\langle N_{ch}\rangle$. However for $\langle N_{ch}\rangle$ smaller than 200 a significant deviation of $\tilde{D}$ from 4 is observed.
It is because the assumptions used to derive the value $\tilde{D}=4$ \cite{Ko.1,Ko.3} are not fulfilled in the limit $\langle N_{ch}\rangle\rightarrow 0$.

\subsection{Model B}

Both measures are constructed in order to be independent of the fluctuations of $N_{ch, tot}$. To test it we varied the total number of charged particles from event to event according to a Gaussian distribution with a mean value 850 and a dispersion of 60 and 120. As in the model A, the global net charge was assumed to be zero.  The results of this simulation are shown in Fig. 2.

\begin{figure}[p]
\epsfig{file=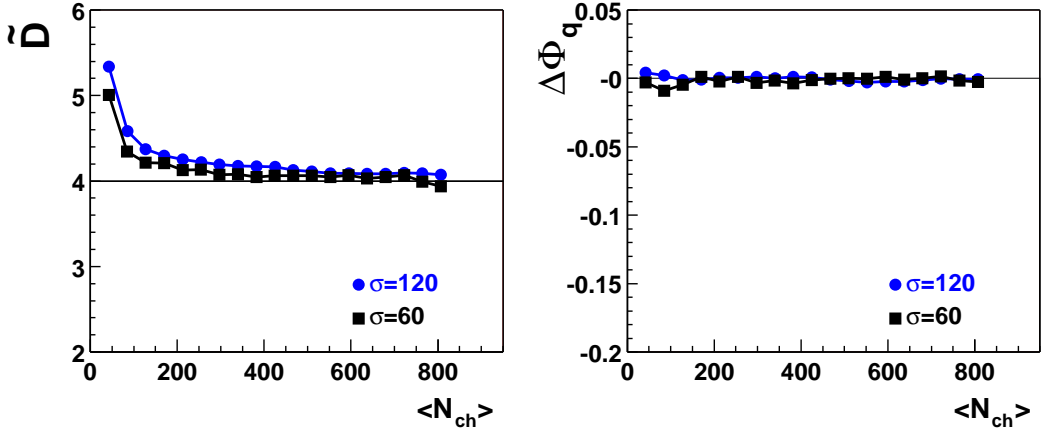,width=14cm}
\caption{
The dependence of $\tilde{D}$ and $\Delta\Phi_q$ on $\langle N_{ch}\rangle$ and $\sigma (N_{ch,tot})$ obtained in model B.
}
\label{fig2}
\end{figure}

\noindent The behaviour of $\tilde{D}$ and $\Delta\Phi_q$ is similar to this obtained in Model A. Thus the influence of multiplicity fluctuations is small, especially on $\Delta\Phi_q$.

\subsection{Model C}

After we have observed that the measures are insensitive to fluctuations of $N_{ch,tot}$ we turn to check the influence of non-zero net charge. We assume $N_{ch,tot}=850$ and $Q=60$. Fig. 3 shows the results.

\begin{figure}[p]
\epsfig{file=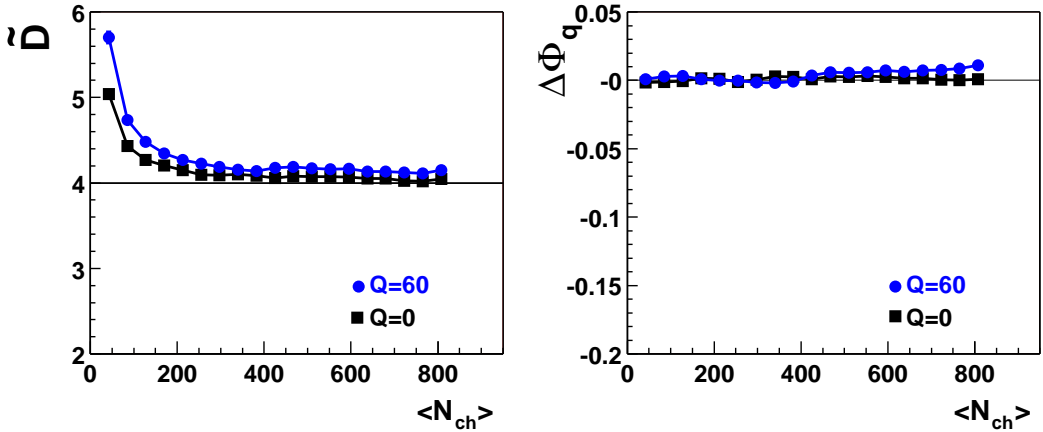,width=14cm}
\caption{
The dependence of $\tilde{D}$ and $\Delta\Phi_q$ on $\langle N_{ch}\rangle$ and net charge $Q$ obtained in model C.
}
\label{fig3}
\end{figure}

\noindent The measure $\tilde{D}$ is explicitly corrected for non-zero net charge by the factor $C_{\mu}$, but it still shows a significant dependence on it. Much weaker dependence on the net charge is observed for $\Delta\Phi_q$. For an acceptance larger than 50\% its value is somewhat higher than the calculated one assuming $Q=0$.

\subsection{Model D}

In nuclear collisions the number of charged particles fluctuates and the net charge is approximately proportional to $N_{ch,tot}$. To see how the measures behave under these assumptions we used a model in which $N_{ch,tot}$ was generated from a Gaussian distribution ($\langle N_{ch,tot}\rangle =850$, $\sigma (N_{ch,tot})=80$) and $N_{pos, tot}=\frac{7}{12}\cdot N_{ch, tot}$,   $N_{neg, tot}=\frac{5}{12}\cdot N_{ch, tot}$ . The results of this simulation are presented in Fig. 4.

\begin{figure}[p]
\epsfig{file=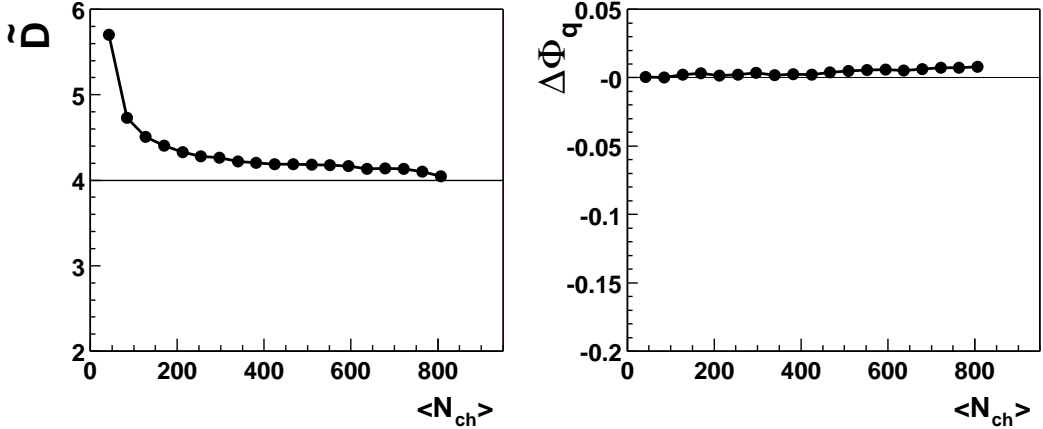,width=14cm}
\caption{
The dependence of $\tilde{D}$ and $\Delta\Phi_q$ on $\langle N_{ch}\rangle$, for fluctuating $N_{ch,tot}$ and for net charge which is proportional to $N_{ch,tot}$ obtained in model D.
}
\label{fig4}
\end{figure}

\noindent The value of $\tilde{D}$ is significantly higher than 4 in the full range of $\langle N_{ch}\rangle$. $\Delta\Phi_q$ is much less dependent on the studied effects. It deviates slightly from 0 for $\langle N_{ch}\rangle >0.5\cdot \langle N_{ch,tot}\rangle$. 

\subsection{QGP Model}

Following Ref. \cite{Ko.1} we construct a simple toy model of the QGP. Under the assumption of zero baryo-chemical potential and providing zero quark mass we calculate the ratio of up-, down-, antiup- and antidown- quarks and gluons in equilibrium. Assuming entropy and net charge conservation in each rapidity window during the evolution from the QGP to the final hadron state the number of pions ($N$) and their net charge is calculated. The number of charged pions is taken to be $N_{ch}=\frac{2}{3}\cdot N$ based on isospin symmetry. Using this model we obtain the results shown in Fig. 5.

\begin{figure}\epsfig{file=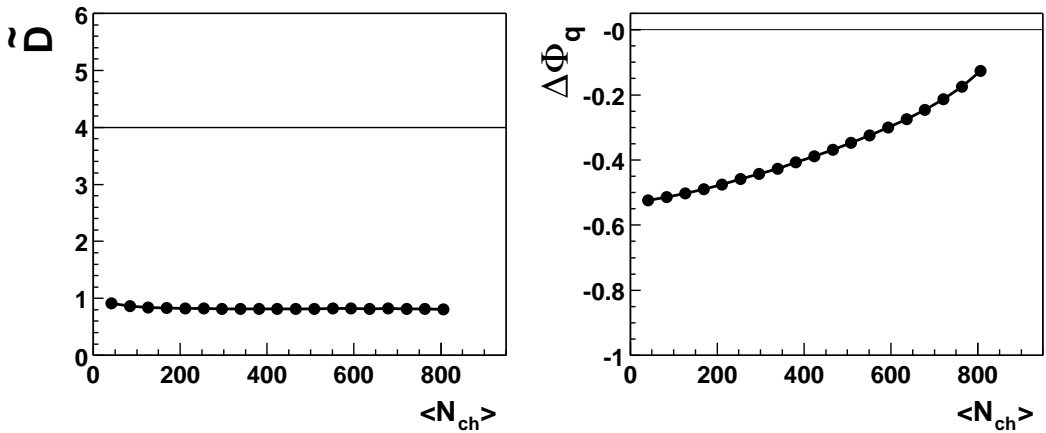,width=14cm}
\caption{
The dependence of $\tilde{D}$ and $\Delta\Phi_q$ on $\langle N_{ch}\rangle$ obtained in a QGP toy model.
}
\label{fig5}
\end{figure}

\noindent Our simulation confirms results published in Ref. \cite{Ko.1}, the $\tilde{D}$ value in our QGP toy model is close to $1$. The new result is the calculation of $\Delta\Phi_q$ in this model. Its value is mostly significantly smaller than zero. It increases from $-0.5$ for small $\langle N_{ch}\rangle$ to $-0.1$ for the largest multiplicity. In case all particles are used, there is no net charge fluctuations due to charge conservation and therefore the value of $\Delta\Phi_q$ should be $0$. We note that the previously discussed trivial effects were increasing $\tilde{D}$ and $\Delta\Phi_q$ to values above $4$ and $0$, respectively. Thus they can not mimic suppression of fluctuations observed in the QGP toy model.

\subsection{'$\rho$'--gas Model}

In this model we study the influence of the resonance decay kinematics on $\tilde{D}$ and $\Delta\Phi_q$. 
We use a MC simulation which generates 400 neutral resonances ('$\rho$') in each event. The mass of the '$\rho$' is fixed to 770 MeV/$c^{2}$. The '$\rho$' rapidity distribution is Gaussian with $\sigma (y)=0.8, 1, 2$ or $3$ and the transverse mass distribution is 'thermal': $\frac{d^2n}{dm_T\cdot dy}=C\cdot m_T\cdot e^{-m_T/T}$ with $T=170$ MeV. All '$\rho$'s decay into two charged pions which are then used to calculate values of $\tilde{D}$ and $\Delta\Phi_q$.
In Fig. 6 we show the results of a simulation for $\sigma(\rm{y}) = 1.0$.  

\begin{figure}[p]
\epsfig{file=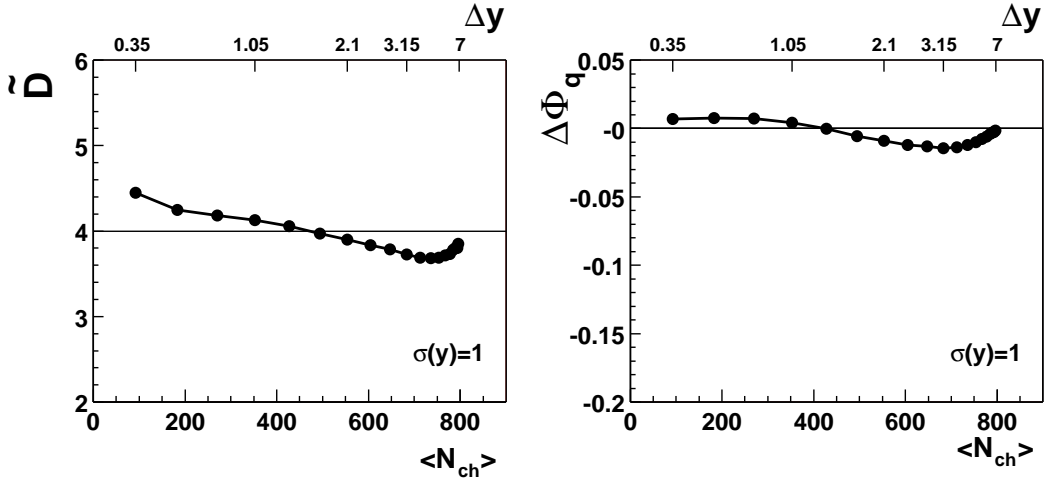,width=14cm}
\caption{
Dependence of  of $\tilde{D}$ and $\Delta\Phi_q$ on $\langle N_{ch}\rangle$ obtained in  '$\rho$'--gas model for $\sigma(\rm{y})=1.0$.
}
\label{fig6}
\end{figure}

\noindent For low $\langle N_{ch}\rangle$ the values of $\tilde{D}$ and $\Delta\Phi_q$ are higher then 4 and 0, respectively. With increasing $\langle N_{ch}\rangle$ these values decrease. 
Resonance decay kinematics influences charge fluctuations in two different ways. It dilutes the effect of global charge conservation if only one of the decay products falls into the acceptance. This will increase the values of $\tilde{D}$ and $\Delta\Phi_q$. If both decay products fall into the acceptance $\langle N_{ch}\rangle$ increases but the net charge doesn't change. $\tilde{D}$ and $\Delta\Phi_q$ measure charge fluctuations normalised by the number of charged particles. Therefore their values will drop in this case. 
Which effect dominates depends on the size of acceptance window ($\Delta \rm{y}$, upper scale on the plots), mass of the decaying resonances and their rapidity distribution. The sensitivity on the last one is shown in Fig. 7. 

\begin{figure}

\epsfig{file=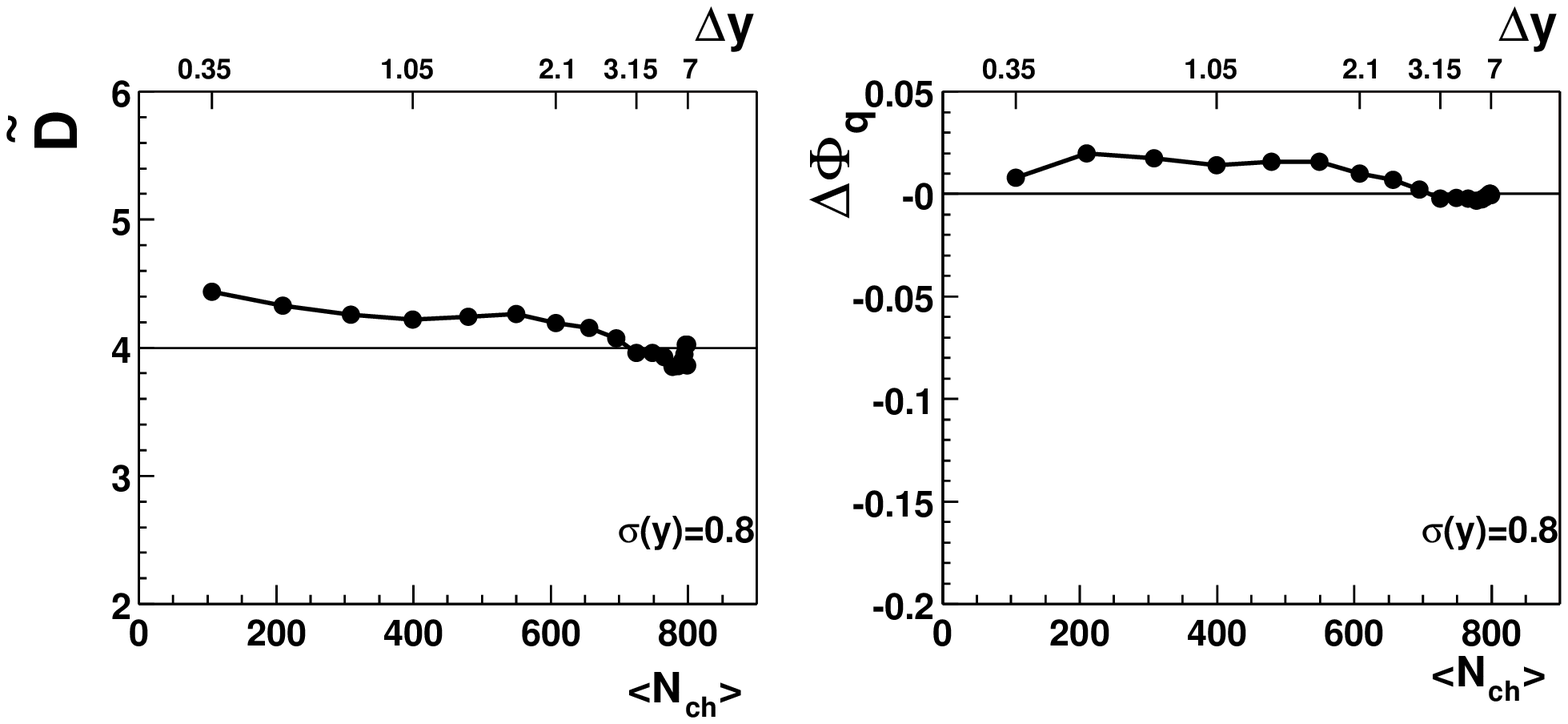,width=13.5cm}

\epsfig{file=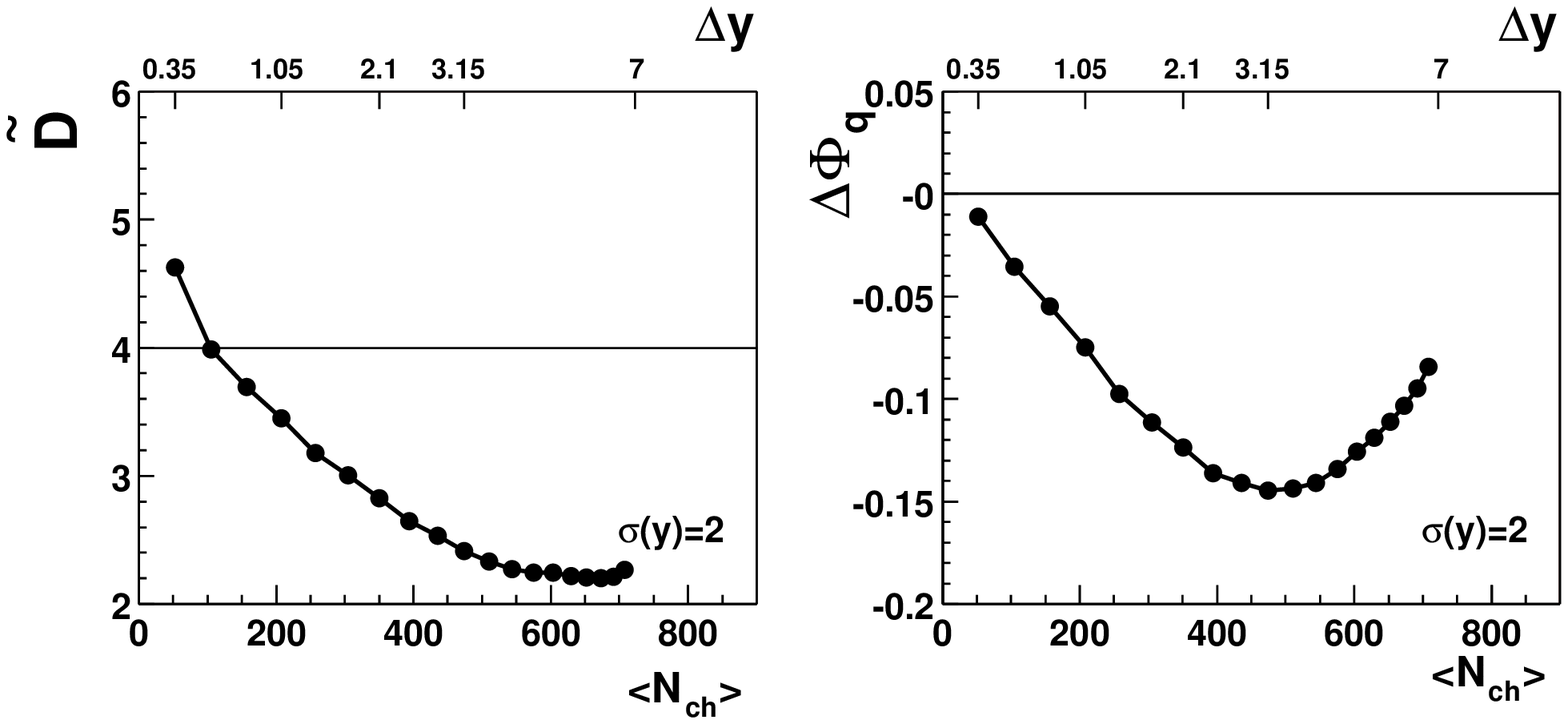,width=13.5cm}

\epsfig{file=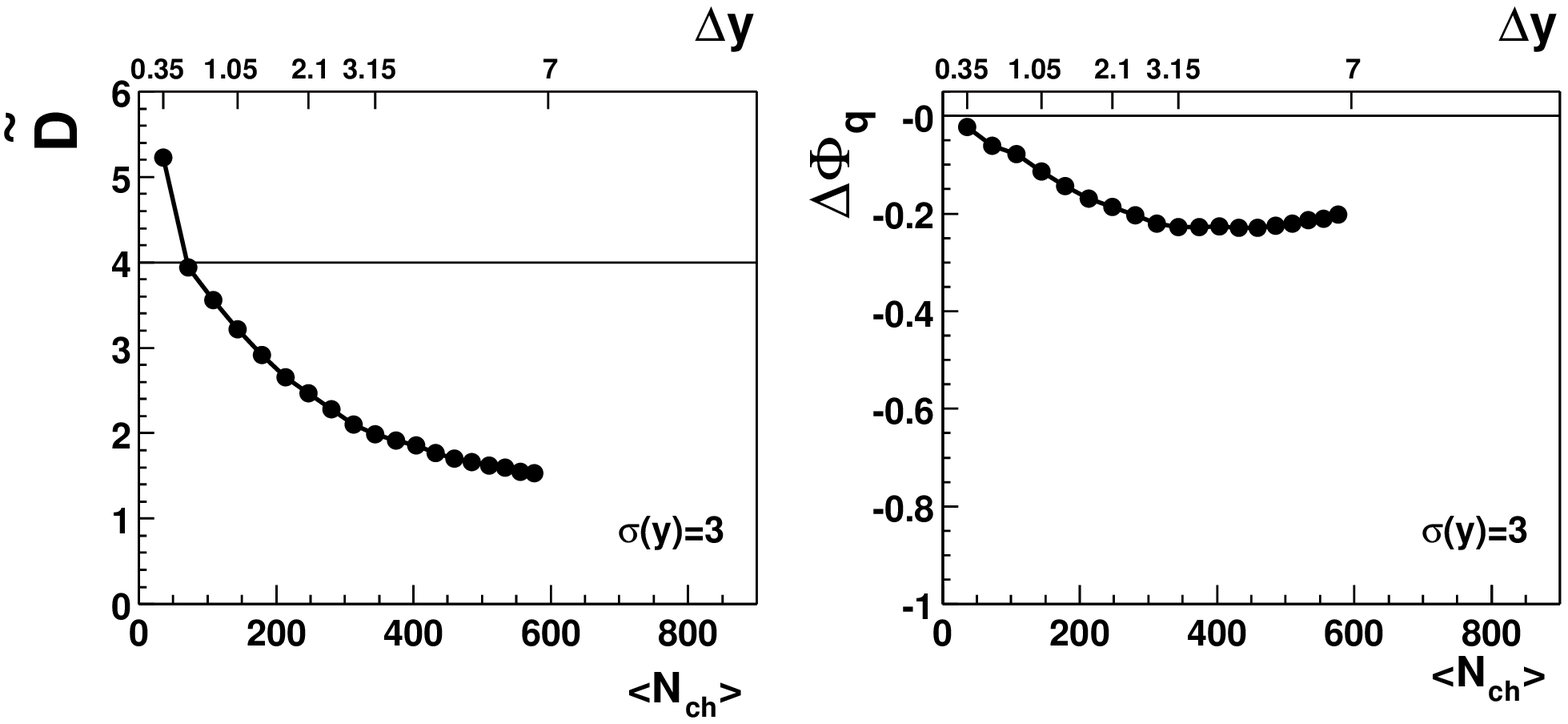,width=13.5cm}
\caption{
Dependence of $\tilde{D}$ and $\Delta\Phi_q$ on $\langle N_{ch}\rangle$ obtained in '$\rho$'--gas model for $\sigma(\rm{y})=0.8, 2.0$ and $3$. Note that horizontal upper scale ($\Delta\rm{y}$) changes with $\sigma(\rm{y})$. 
}
\label{fig7}
\end{figure}

\noindent At SPS energies $\sigma(\rm{y})$ of the $\rho$ distribution is expected to be close to $1$ \cite{Na49.1}. In Figs. 6 and 7 (upper plots) we see that small changes of $\sigma(\rm{y})$ around $1$ have strong influence on the values of $\tilde{D}$ and $\Delta\Phi_q$. For $\sigma(\rm{y})=1$ we observe a transition from fluctuation enhancement to suppression with an increasing rapidity window. For $\sigma(\rm{y})=0.8$ the values of $\tilde{D}$ and $\Delta\Phi_q$ are above or equal to their values expected for independent particle production with global charge conservation for all $\Delta \rm{y}$. 
We would like to stress here that in our model the '$\rho$'s are neutral and therefore there are no charge fluctuations before the '$\rho$'s decay. In Fig. 6 and Fig. 7 (upper plots) we see that for $\sigma(\rm{y})$ values expected at CERN SPS the resonance decay kinematics increases the initial zero charge fluctuations to values which are close to 4 or 0, respectively. 
The fluctuations are significantly lower for larger values of $\sigma(\rm{y})$, e.g. $\sigma(\rm{y})=2$ and $3$, as are expected at RHIC and LHC energies. In the case of QGP creation the measured fluctuations will be composed of the initial QGP fluctuations and fluctuations produced by the decay of resonances. The latter ones are, as our model shows, still larger than the predicted QGP fluctuations even for $\sigma(\rm{y})=2$ and $3$. Therefore we conclude that resonance decay kinematics may significantly shadow charge fluctuations developed at the early stage of the collision.

\section{Summary and Conclusions}

The properties of two measures of charge fluctuations $\tilde{D}$ and $\Delta\Phi_q$ were studied within several models. 
We have shown the dependence of both measures on trivial effects, in particular, on number of particles used per event, on the net electric charge and on fluctuations in the total event multiplicity. Furthermore their response to resonances decay kinematics and to the reduction of fluctuations due to QGP creation were analysed, too.
From our simulations we learn that the measure $\Delta\Phi_q$ is significantly less dependent on trivial effects than $\tilde{D}$ and that both measures are sensitive to frozen QGP fluctuations and to resonance decay kinematics. Therefore we suggest to use $\Delta\Phi_q$ as a proper measure of charge fluctuations. Using the '$\rho$' gas model we have shown that the influence of resonance decay strongly depends on their rapidity distribution and on the acceptance window. 
Given the typical widths of the rapidity distributions of $\rho$'s at CERN SPS energies they can easily increase charge fluctuations from QGP values to values expected for independent particle production with global charge conservation. A further development of methods which allow to distinguish various sources of charge fluctuations is obviously necessary.

\vspace{1cm}
\noindent
{\bf Acknowledgements}

I am very indebted to Marek Ga\'zdzicki for numerous fruitful discussions and ideas. I'm also grateful to S. Mr\'owczy\'nski, V. Koch, H. Str\"obele, R. Stock, P. Seyboth, R. Renfordt and J. Berschin for helpful comments.

\end{document}